\documentclass[aps,prc,twocolumn,showpacs,preprintnumbers,amsmath,amssymb]{revtex4}
\usepackage{graphicx}
\usepackage{dcolumn}
\usepackage{bm}
\usepackage{longtable}

\begin{document}

\title{Dipole and Quadrupole Electroexcitations of the Isovector $T=1$ Particle-Hole States in $^{12}$C\,\,\footnote{\,\,Support by Conselho Nacional de Desenvolvimento Cient\'{\i}fico e
Tecnol\'{o}gico (CNPq) (Brazil), and the Third World Academy of
Science (TWAS) (Italy) for under grant of the scheme (TWAS-CNPq
exchange programs for postdoctoral researchers).}}

\author{F.~A.~Majeed$^{1,\footnote{\,\,Permenant address: Department of Physics, College of Science, Al-Nahrain University, Baghdad, IRAQ. \,\,\,Email:fouad@if.ufrj.br}}$, R.~A.~Radhi$^2$}

\address{$^1$Instituto de F\'{\i}sica, Universidade Federal do
Rio de Janeiro, C.P. 68528, 21941-972 Rio de Janeiro, RJ, Brazil}
\address{$^2$Department of Physics, College of Science, Baghdad
University, Baghdad, IRAQ}

\date{\today}

\begin{abstract}
Electroexcitations of the dominantly $T=1$ particle-hole states of
$^{12}$C are studied in the framework of the harmonic oscillator
shell model. All possible $T=1$ single-particle-hole states of all
allowed angular momenta are considered in a basis including
single-particle states up to the 1$f$-2$p$ shell. The Hamiltonian is
diagnoalized in this space in the presence of the modified surface
delta interaction. Correlation in the ground state wave functions by
mixing more than one configuration is considered and shows a major
contribution that leads to enhance the calculations of the form
factors. A comparison with the experiment shows that this model is
able to fit the location of states and a simple scaling of the
results give a good fit to the experimental form factors.
\end{abstract}

\maketitle

Nuclear structure models can be successfully tested by comparing the
calculated and measured electron scattering form factors. The
success of such a model reveals a valuable information about the
charge and current distributions of nuclei.

The transverse electric form factor of the 2$^{+}$ level at 16.107
MeV $T=1$ of $^{12}$C were measured by Flanz {\em et
al.,}\cite{FB78} by means of 180$^{\circ}$ electron scattering over
a momentum-transfer range from $q$=0.51 to 2.05 fm$^{-1}$.
Deutschmann {\em et al.,}\cite{DU83} studied the inelastic electron
scattering cross section for the M1 transition to the 15.11 MeV
(1$^{+}$,$T=1$) level and for M2 transition to the 16.58 MeV
(2$^{-}$,$T=1$) level in $^{12}$C in the momentum-transfer region
($q$=0.4--3.0 fm$^{-1}$). Electron scattering at 200 MeV on $^{12}$C
and $^{13}$C, have been studied by T. Sato {\em et al.,} \cite{ST85}
to investigate magnetic dipole M1 and electric quadrupole E2 nuclear
form factors. The effect of higher configurations wave functions are
investigated by Bennhold {\em et al.,} \cite{BC92}. Donnelly
\cite{DT70} studied the $T=1$ single-particle-hole states of
$^{12}$C on the basis of the harmonic oscillator (HO) shell model in
the particle-hole formalism developed by Lewis and Walecka,
configuration mixing is included via a Serber--Yukawa residual
interaction.

In this Letter, we study the isovector ($T=1$) transition in
$^{12}$C which connects the ($J$$^{\pi}$ =0$^{+}$, $T$=0) ground
state with the ($J$$^{\pi}$ =2$^{-}$, 3$^{-}$ and 3$^{+}$) isovector
states. The ground state is taken to have closed 1$s$$_{1/2}$ and
1$p$$_{1/2}$ shells. The states expected to be most strongly excited
from closed-shell nuclei are linearly combination of a
configurations in which one nucleon has been raised to a higher
shell, forming pure single-particle-hole state \cite{TW84}. This
approximation is called Tamm-Dancoff approximation (TDA)\cite{TJ66}.
The dominant dipole and quadrupole $T=1$ single particle-hole states
of $^{12}$C are considered with the framework of the harmonic
oscillator (HO) shell model. The Hamiltonian is diagnoalized in the
space of the single-particle hole states, in the presence of the
modified surface delta interaction (MSDI) \cite{PM77}. The space of
the single-particle-hole states include all shells up to
2$p$$_{1/2}$ shell. Admixture of higher configurations is also
considered. A comparison of the calculated form factors using this
model with the available experimental data for the dominantly $T=1$
states are discussed.

The ground state of $^{12}$C is taken to have closed 1$s$$_{1/2}$
and 1$p$$_{3/2}$ shells, and is represented by $\Psi$$_{0}$. A state
formed by the promotion of one particle from the shell-model ground
state is called a particle-hole state. The particle-hole state of
the total Hamiltonian is represented by $\Phi$$_{JM}$($ab$$^{-1}$)
with labels (a) for particles with quantum numbers
($n$$_{a}$$\ell$$_{a}$$j$$_{a}$) and (b) for holes with quantum
numbers ($n$$_{b}$$\ell$$_{b}$$j$$_{b}$). The state
$\Phi$$_{JM}$($ab$$^{-1}$) indicating that a particle was vacated
from $j$$_{b}$ and promoted to $j$$_{a}$.

The excited state wave function can be constructed as a linear
combinations of pure basis  $\Phi$$^{,s}$  as \cite{TW84}
\begin{equation}
\Psi^{n}_{JM}=\sum_{ab}\chi^{J}_{ab^{-1}}\Phi_{JM}(ab^{-1}),
\end{equation}
where the amplitude $\chi$$^{J}_{ab^{-1}}$ can be determined from a
diagonalization of the residual interaction. By including the
isospin $T$, \cite{TJ66} one now has to solve the secular equation
\begin{equation}
\sum_{ab}[\langle{\acute{a}\acute{b}^{-1}}|H|ab^{-1}\rangle_{JMTT_z}-E_n\delta_{\acute{a}\acute{b}^{-1},
ab^{-1}}]\,\chi^{JT}_{ab^{-1}}=0.
\end{equation}
The matrix element of the Hamiltonian is given by \cite{PM77}
\begin{eqnarray}
\langle{\acute{a}\acute{b}^{-1}}|H|ab^{-1}\rangle_{JMTT_z}
=(e_{\acute{a}}-e_{\acute{b}})\,\delta_{{a\acute{a}},{b\acute{b}}}\nonumber\\
+\langle{\acute{a}\acute{b}^{-1}}|V|ab^{-1}\rangle_{JMTT_z},
\end{eqnarray}
where $e$$_{\acute{a}}$-$e$$_{\acute{b}}$ is the unperturbed energy
of the particle-hole pair obtained from energies in nuclei with
A$\pm$1 particles.

The matrix element of the residual interaction $V$ is given by the
modified surface delta interaction (MSDI) with the strength
parameters $A$$_{0}$=0.8 MeV, $A$$_{1}$=1.0 MeV, $B$=0.7 MeV and
$C$=$-$0.3 MeV \cite{PM77}
\begin{eqnarray}
\langle{\acute{a}\acute{b}^{-1}}|V|ab^{-1}\rangle_{JMTT_z}
=-\sum_{\acute{J}\acute{T}}(2\acute{J}+1)(2\acute{T}+1)\nonumber\\\times\left\{
\begin{array}{ccc}
j_{\acute{a}} &  j_{b} & \acute{J}\\
j_{a} & j_{\acute{b}} & J
\end{array}\right\}\left\{
\begin{array}{ccc}
\frac{1}{2} &  \frac{1}{2} & T\\
\frac{1}{2} & \frac{1}{2} & \acute{T}
\end{array}\right\}\langle{\acute{a}\,b}|V|a\acute{b}\rangle_{\acute{J}\,\acute{T}}.
\end{eqnarray}

The matrix elements of the multipole operators $T$$_{J}$ are given
in terms of the single particle matrix elements by \cite{TW84}
\begin{equation}
     \left\langle\Psi_{J}\|T_{Jt_{z}}\|\Psi_{0}\right\rangle
     =\sum_{ab}\chi^{Jt_{z}}_{ab^{-1}}\left\langle
     a\|T_{Jt_{z}}\|b\right\rangle,
\end{equation}
where $t$$_{z}$=1/2 for protons and -1/2  for neutrons. The
amplitudes $\chi$$^{Jt_{z}}_{ab^{-1}}$ can be written in terms of
the amplitudes $\chi$$^{JT}_{ab^{-1}}$ in isospin space as
\cite{PM77}
\begin{eqnarray}\
      \chi^{Jt_{z}}_{ab^{-1}}=(-1)^{T_{f}-T_{i}}\left[\left(
                                                        \begin{array}{ccc}
                                                          T_{f} & 0 & T_{i} \\
                                                          -T_{z} & 0 & T_{z} \\
                                                        \end{array}
                                                      \right)\sqrt{2}\ \frac{\chi^{JT=0}_{ab^{-1}}}{2}\right.\nonumber\\
      \left.+2t_{z}\left(\begin{array}{ccc}
                  T_{f} & 0 & T_{i} \\
                  -T_{z} & 0 & T_{z} \\
                  \end{array}
                  \right)\sqrt{6}\
                  \frac{\chi^{JT=1}_{ab^{-1}}}{2}\right],
\end{eqnarray}
where
\begin{equation}\
    T_{z}=\frac{Z-N}{2}
\end{equation}

The single particle matrix elements of the electron scattering
operator $T$$^{\eta}_{J}$ are those of Ref.\cite{BA85} with $\eta$
selects the longitudinal ($L$), transverse electric ($E\ell$) and
transverse magnetic ($M$) operators, respectively. Electron
scattering form factors involving angular momentum transfer $J$ is
given by \cite{BA85}
\begin{eqnarray}\
    |F^{\eta}_{J}(q)|^{2}=\frac{4\pi}{Z^{2}(2J_{i}+1)}\ |\langle\Psi_{J_{f}}\|T^{\eta}_{Jt_{z}}\|\Psi_{J_{i}}\rangle\nonumber\\
    \times|F_{c.m}(q)|^{2} \ |F_{f.s}(q)|^{2}
\end{eqnarray}
where $J$$_{i}$= 0 and $J$$_{f}$=$J$ for closed shell nuclei and $q$
is the momentum transfer. The last two terms in Eq.\,(8) are the
correction factors for the c.m. and the finite nucleon size
($f.s.$)\cite{BA85}. The total inelastic electron scattering form
factor is defined as \cite{TJ66}
\begin{equation}
     |F_{J}(q,\theta)|^{2}=|F^{L}_{J}(q)|^{2}+\left[\frac{1}{2}+\tan^{2}(\theta/{2})\right]
     |F^{Tr}_{J}(q)|^{2},
\end{equation}
where $|F^{Tr}_{J}(q)|^{2}$ is the transverse electric or transverse
magnetic form factors.

The resulting particle-hole states are listed in Table~\ref{tab1}
and Table~\ref{tab2} together with the values of $J^{\pi}$, for the
positive and negative parity states, respectively. The configuration
energies $e_{a}-e_{b}$ of these states obtained from the spectra of
$^{13}$C and  $^{11}$C \cite{UD83} are also given in these tables.
\begingroup
\begin{table}
 \caption{\label{tab1} The unperturbed energies of the particle-hole
 positive parity states  and  the possible $J$ and $T=1$ values in  $^{12}$C.}
\begin{ruledtabular}
\begin{center}
\begin{tabular}{ccl}

  Particle-hole &  Particle-hole \\
  configuration &  configuration energies & $J$\\
  $|$a\,$b^{-1}$$\rangle$ & ($e_{a}-e_{b}$)\,MeV\\
\hline \\
 (1$p$$_{1/2}$)\,(1$p$$_{3/2}$)$^{-1}$ & 13.77 & 1, 2       \\
 (1$d$$_{5/2}$)\,(1$s$$_{1/2}$)$^{-1}$ & 33.90 & 2, 3       \\
 (2$s$$_{1/2}$)\,(1$s$$_{1/2}$)$^{-1}$ & 33.14 & 0, 1       \\
 (1$d$$_{1/2}$)\,(1$s$$_{1/2}$)$^{-1}$ & 38.38 & 1, 2       \\
 (1$f$$_{7/2}$)\,(1$p$$_{3/2}$)$^{-1}$ & 25.74 & 2, 3, 4, 5 \\
 (2$p$$_{3/2}$)\,(1$p$$_{3/2}$)$^{-1}$ & 27.37 & 0, 1, 2, 3 \\
 (1$f$$_{5/2}$)\,(1$p$$_{3/2}$)$^{-1}$ & 34.17 & 1, 2, 3, 4 \\
 (2$p$$_{1/2}$)\,(1$p$$_{3/2}$)$^{-1}$ & 33.37 & 1, 2       \\
\end{tabular}
\end{center}
\end{ruledtabular}
\end{table}
\endgroup

\begingroup
\begin{table}
 \caption{\label{tab2} The unperturbed energies of the particle-hole
 negative parity states  and  the possible $J$ and $T=1$ values in  $^{12}$C.}
\begin{ruledtabular}
\begin{center}
\begin{tabular}{ccl}

  Particle-hole &  Particle-hole \\
  configuration &  configuration energies & $J$\\
  $|$a\,$b^{-1}$$\rangle$ & ($e_{a}-e_{b}$)\,MeV\\
\hline \\
 (1$p$$_{1/2}$)\,(1$s$$_{1/2}$)$^{-1}$ & 30.05 & 0, 1       \\
 (1$d$$_{5/2}$)\,(1$p$$_{3/2}$)$^{-1}$ & 17.62 & 1, 2, 3, 4 \\
 (1$d$$_{3/2}$)\,(1$p$$_{3/2}$)$^{-1}$ & 22.11 & 0, 1, 2, 3 \\
 (2$s$$_{1/2}$)\,(1$p$$_{3/2}$)$^{-1}$ & 16.86 & 1, 2       \\
 (1$f$$_{7/2}$)\,(1$s$$_{1/2}$)$^{-1}$ & 42.02 & 3, 4       \\
 (2$p$$_{3/2}$)\,(1$s$$_{1/2}$)$^{-1}$ & 43.65 & 1, 2       \\
 (1$f$$_{5/2}$)\,(1$s$$_{1/2}$)$^{-1}$ & 50.45 & 2, 3       \\
 (2$p$$_{1/2}$)\,(1$s$$_{1/2}$)$^{-1}$ & 49.65 & 0, 1       \\
\end{tabular}
\end{center}
\end{ruledtabular}
\end{table}
\endgroup
Higher configurations are included in the calculations when the
ground state is considered as a mixture of the
$|(1$$s$$_{1/2}$)$^{4}$$\,(1$$p$$_{3/2}$)$^{8}$$\rangle$ and
$|(2$$s$$_{1/2}$)$^{4}$$\,(2$$p$$_{3/2}$)$^{8}$$\rangle$
configurations, such that the ground state wave function becomes
\begin{eqnarray}
|\Psi_{00}\rangle=\gamma|\Psi_{00}(1s_{1/2})^4(1p_{3/2})^8\rangle\nonumber\\
                                        +\delta|\Psi_{00}(2s_{1/2})^4(2p_{3/2})^8\rangle
\end{eqnarray}

with $\gamma^{2}$+$\delta^{2}$=1,
$\chi^{JT}_{ab_{1}^{-1}}$=$\gamma\chi^{JT}_{ab^{-1}}$ and
$\chi^{JT}_{ab_{2}^{-1}}$=$\delta\chi^{JT}_{ab^{-1}}$

The excited states is also assumed as a mixture of the particle-hole
configurations, $|$a$_{1}$\,$b^{-1}_{1}$$\rangle$,
$|$a$_{2}$\,$b^{-1}_{2}$$\rangle$, $|$a$_{2}$\,$b^{-1}_{1}$$\rangle$
and $|$a$_{1}$\,$b^{-1}_{2}$$\rangle$, where
$|$a$_{1}$$\rangle$=$|$a$\rangle$=$|$$n$$_{a}$$\,$$\ell$$_{a}$$\,$$j$$_{a}$$\rangle$,
$|$a$_{2}$$\rangle$=$|$a$\rangle$=$|$$n$$_{a}$$+1\,$$\ell$$_{a}$$\,$$j$$_{a}$$\rangle$,
$|$b$_{1}$$\rangle$=$|$b$\rangle$=$|$$n$$_{b}$$\,$$\ell$$_{b}$$\,$$j$$_{b}$$\rangle$
and
$|$b$_{2}$$\rangle$=$|$b$\rangle$=$|$$n$$_{b}$$+1\,$$\ell$$_{b}$$\,$$j$$_{b}$$\rangle$.
The configurations which include the higher configurations is called
the extended space configurations.

The matrix element given in Eq.\,(5) is called the model space
matrix element, where $a$ and $b$ are defined by the amplitudes
given in Table~\ref{tab3} and Table~\ref{tab4}, respectively. The
extended space matrix element becomes

\begin{eqnarray}
     \left\langle\Psi_{J}\|T_{Jt_{z}}\|\Psi_{0}\right\rangle
     =\sum_{a_{1}b_{1}}\chi^{Jt_{z}}_{a_{1}b_{1}^{-1}}\left\langle
     a_{1}\|T_{Jt_{z}}\|b_{1}\right\rangle\nonumber\\
     +\sum_{a_{1}b_{2}}\chi^{Jt_{z}}_{a_{1}b_{2}^{-1}}\left\langle
     a_{1}\|T_{Jt_{z}}\|b_{2}\right\rangle\nonumber\\
     +\sum_{a_{2}b_{1}}\chi^{Jt_{z}}_{a_{2}b_{1}^{-1}}\left\langle
     a_{2}\|T_{Jt_{z}}\|b_{1}\right\rangle\nonumber\\
     +\sum_{a_{2}b_{2}}\chi^{Jt_{z}}_{a_{2}b_{2}^{-1}}\left\langle
     a_{2}\|T_{Jt_{z}}\|b_{2}\right\rangle,
\end{eqnarray}
where
\begin{eqnarray}
    \chi^{Jt_{z}}_{a_{1}b_{1}^{-1}}=C_{1}\,\chi^{Jt_{z}}_{ab^{-1}},\nonumber\\
    \chi^{Jt_{z}}_{a_{1}b_{2}^{-1}}=C_{2}\,\chi^{Jt_{z}}_{ab^{-1}},\nonumber\\
    \chi^{Jt_{z}}_{a_{2}b_{1}^{-1}}=C_{3}\,\chi^{Jt_{z}}_{ab^{-1}},\nonumber\\
    \chi^{Jt_{z}}_{a_{2}b_{2}^{-1}}=C_{4}\,\chi^{Jt_{z}}_{ab^{-1}},
\end{eqnarray}

\begingroup
\begin{table}
 \caption{\label{tab3} Energies and amplitudes $\chi$$^{JT}$ for $J$$^{+}$ $T=1$ states.}
\begin{ruledtabular}
\begin{center}
\begin{tabular}{ccc}

  Particle-hole &  E($1$$^{+}$)=15.54 MeV & E($2$$^{+}$)=15.49 MeV \\
  configuration &  $\chi$$^{11}$ & $\chi$$^{21}$\\
  $|$a\,$b^{-1}$$\rangle$ & \\
\hline \\
 (1$p$$_{1/2}$)\,(1$p$$_{3/2}$)$^{-1}$ & 0.99671 & 0.99774 \\
 (1$d$$_{5/2}$)\,(1$s$$_{1/2}$)$^{-1}$ & 0.00000 & 0.00321 \\
 (2$s$$_{1/2}$)\,(1$s$$_{1/2}$)$^{-1}$ & 0.03900 & 0.00000 \\
 (1$d$$_{1/2}$)\,(1$s$$_{1/2}$)$^{-1}$ & 0.01378 & 0.03724 \\
 (1$f$$_{7/2}$)\,(1$p$$_{3/2}$)$^{-1}$ & 0.00000 & 0.00892 \\
 (2$p$$_{3/2}$)\,(1$p$$_{3/2}$)$^{-1}$ & 0.05340 & -0.01953\\
 (1$f$$_{5/2}$)\,(1$p$$_{3/2}$)$^{-1}$ & 0.01823 & 0.03449 \\
 (2$p$$_{1/2}$)\,(1$p$$_{3/2}$)$^{-1}$ & 0.04104 & 0.03828 \\
\end{tabular}
\end{center}
\end{ruledtabular}
\end{table}
\endgroup

\begingroup
\begin{table}
 \caption{\label{tab4} Energies and amplitudes $\chi$$^{JT}$ for $2$$^{-}$ $T=1$ state.}
\begin{ruledtabular}
\begin{center}
\begin{tabular}{cc}

  Particle-hole &  E($2$$^{-}$)=18.19 MeV \\
  configuration &  $\chi$$^{21}$ \\
  $|$a\,$b^{-1}$$\rangle$ & \\
\hline \\
 (1$p$$_{1/2}$)\,(1$p$$_{3/2}$)$^{-1}$ & 0.00000  \\
 (1$d$$_{5/2}$)\,(1$s$$_{1/2}$)$^{-1}$ & -0.54940 \\
 (2$s$$_{1/2}$)\,(1$s$$_{1/2}$)$^{-1}$ & -0.03147 \\
 (1$d$$_{1/2}$)\,(1$s$$_{1/2}$)$^{-1}$ & 0.83484  \\
 (1$f$$_{7/2}$)\,(1$p$$_{3/2}$)$^{-1}$ & 0.00000  \\
 (2$p$$_{3/2}$)\,(1$p$$_{3/2}$)$^{-1}$ & 0.01049  \\
 (1$f$$_{5/2}$)\,(1$p$$_{3/2}$)$^{-1}$ & 0.02690  \\
 (2$p$$_{1/2}$)\,(1$p$$_{3/2}$)$^{-1}$ & 0.00000  \\
\end{tabular}
\end{center}
\end{ruledtabular}
\end{table}
\endgroup

The values of the parameters C$^{,s}$ are given in Table~\ref{tab5}.
Experimentally the states $1$$^{+}$, $2$$^{+}$ and $2$$^{-}$ are
found at 15.11 MeV, 16.11 MeV and 16.58 MeV, respectively
\cite{RR87}. We obtain the values 15.54 MeV, 15.52 MeV and 18.19 MeV
for the states $1$$^{+}$, $2$$^{+}$ and $2$$^{-}$, respectively.

The $1$$^{+}$ (15.11 MeV), M1 form factor is shown in Fig.\,1. The
dashed curve represents the calculations with the amplitudes
$\chi$$^{,s}$ reduced by a factor 2, to agree with the low $q$ data
\cite{TW84}. The single-particle matrix elements are calculated with
the harmonic oscillator wave functions (HO) with oscillator
parameter $b=1.64$\, fm  to agree with the elastic form factor
determination \cite{TK85}. A best fit to the M1 form factor data is
obtained by Hicks {\em et al.} \cite{RR87} with $b=1.67$\,fm.
\begin{figure}
\centering
\includegraphics[width=0.49\textwidth]{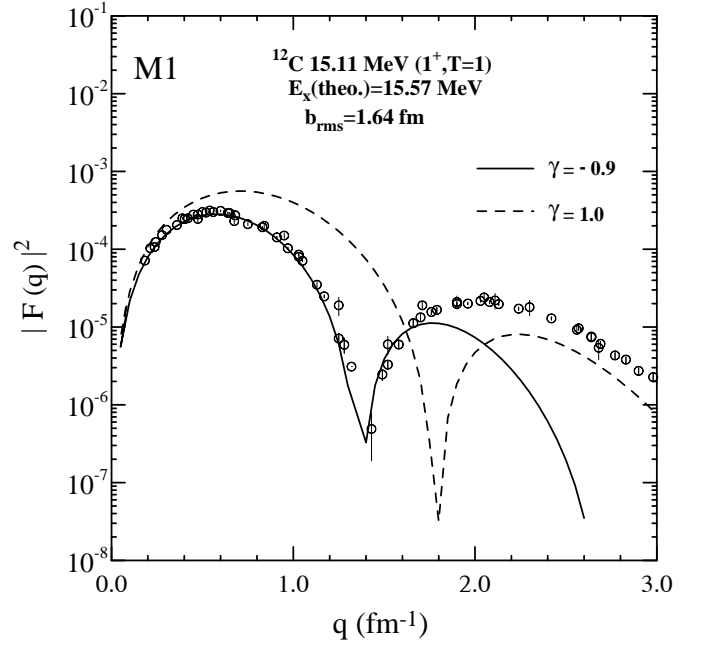}\,
\caption{Form factor for the M1 transition to the ($1$$^{+}$, 1)
15.57 MeV state compared with the experimental data taken from Ref.
\cite{UD83}.}
\end{figure}

Karataglidis {\em et al.} \cite{SK95} adopted the value 1.7\,fm in
describing the form factors for the  $0\rightarrow 2^{+}$ (4.44 MeV
and 16.11 MeV) transitions in $^{12}$C. The low momentum transfer
data for the 15.11 MeV $1$$^{+}$ state are fitted with $b=1.9$\,fm
\cite{FG64}.

This $b$ value is adopted also by Donnelly \cite{TW84}, to describe
the M1 form factors up to momentum transfer 1.4 fm$^{-1}$. In our
work we fix the value of  $b=1.64$\,fm and trying to find other
contributions that can describe the experimental form factors. When
the M1 form factor is calculated with the extended space
configurations, the result is shown by the solid curve in Fig.\,1.
The data are very well described up to momentum-transfer
1.7\,fm$^{-1}$ , and the form factor is shifted to agree with the
location of the experimental diffraction minimum. The calculated
form factor decreases too steeply compared to the data at high q
values. All previous theoretical treatments shows this behaviour.
Our results are consistent with those of  Sato {\em et al.}
\cite{TK85}, where the meson exchange current is included but shows
a minor contribution.

The Coulomb C2 and the transverse E2 form factors are shown in Figs.
\,2 and 3. The amplitudes $\chi$ have to be reduced also by a factor
2 to fit the low $q$ data. The coulomb form factor is calculated
with core-polarization effects by introducing effective charges
\cite{BA83} for the protons (1.25$e$) and for the neutrons (0.5$e$).
With these effective charges, the experimental data are very well
described for the region of the momentum transfer $q \leq$ 2.2
fm$^{-1}$. The same agreement are obtained for the E2 form factor.
\begin{figure}
\centering
\includegraphics[width=0.49\textwidth]{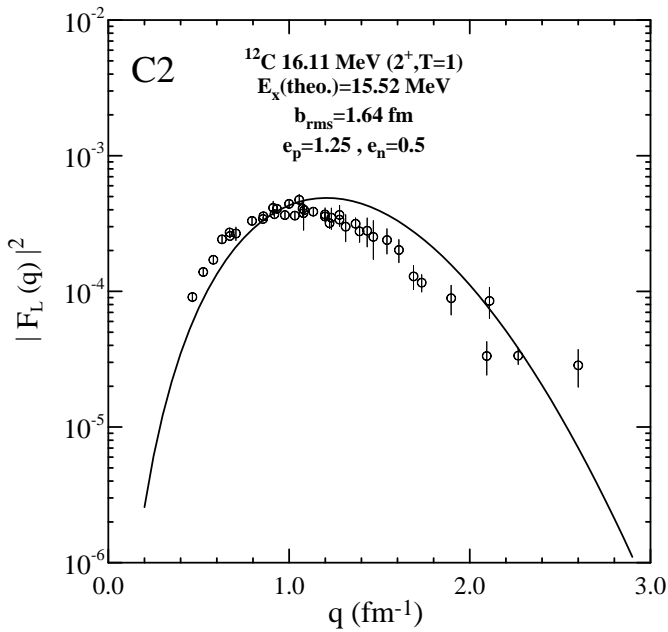}
\caption{Longitudinal form factor for the C2 transition to the
(2$^{+}$, 1) 15.52 MeV state compared with the experimental data
taken from Ref. \cite{LB89}.}
\end{figure}

\begin{figure}
\centering
\includegraphics[width=0.48\textwidth]{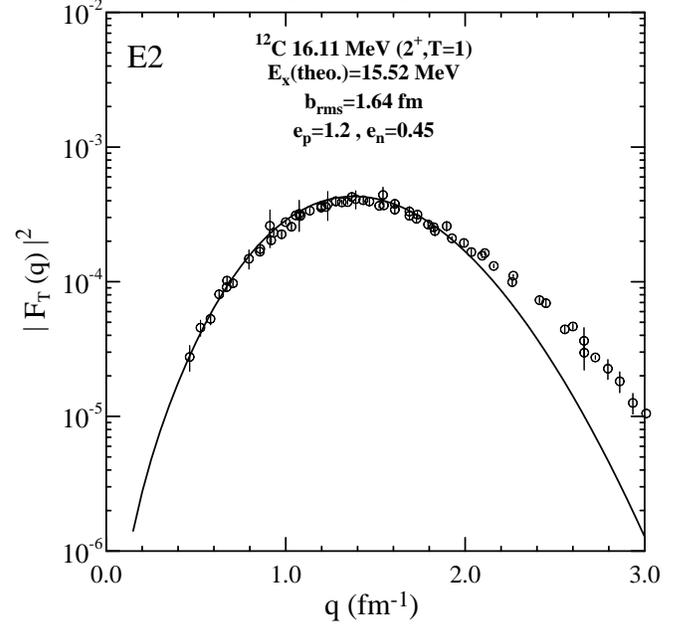}
\caption{Transverse electric form factor for the E2 transition to
the (2$^{+}$, 1) 15.52 MeV state compared with the experimental data
taken from Ref. \cite{LB89}.}
\end{figure}
The transverse M2 form factor for the excitation to the 2$^{-}$
16.58 MeV state are shown in Fig.\,4.~The dashed curve represents
the model space calculations with $b=1.64$\,fm, where the data can
not be satisfactorily described, especially in the region of the low
$q$ values. Due to the absence of an accepted model, Hicks {\em et
al.} \cite{RR87} used a simple phenomenological procedure to
interpret the data. In our work the data are well described using
the extended space configurations and $b=1.35$\,fm. The amplitudes
are quenched by 86\%. The results are consistent with those of Hicks
{\em et al.,}\cite{RR87} a diffraction minimum is obtained at $q
\cong$ 0.6 fm$^{-1}$.

\begin{figure}
\centering
\includegraphics[width=0.48\textwidth]{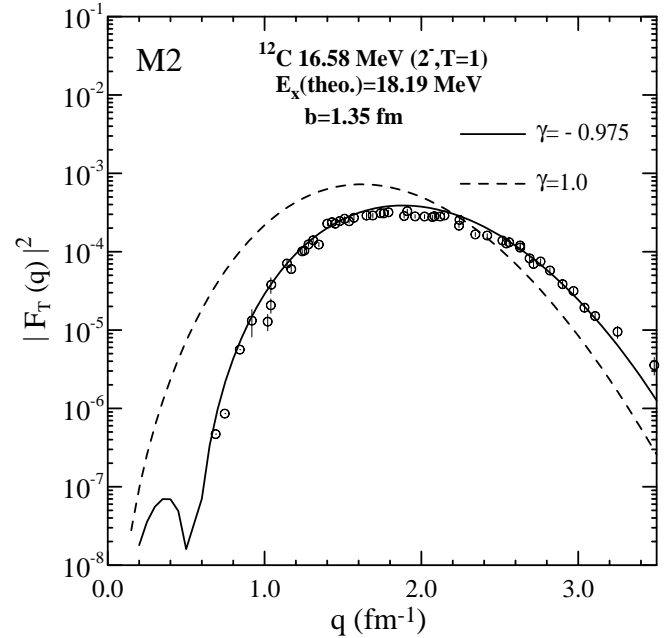}
\caption{Transverse magnetic form factor for the M2 transition to
the (2$^{-}$, 1) 18.19 MeV state compared with the experimental data
taken from Ref. \cite{UD83}.}
\end{figure}

\begingroup
\begin{table}
 \caption{\label{tab5} Values of the parameters $C$$^{,s}$  used in the extended space calculations.}
\begin{ruledtabular}
\begin{center}
\begin{tabular}{ccccc}

  $J$$^{\pi}$ &  $C$$_{1}$ & $C$$_{2}$ & $C$$_{3}$ & $C$$_{4}$\\
 \hline \\
 1$^{+}$ & 0.92 & -0.27 & -0.27 & 0.078 \\
 2$^{-}$ &  -0.92 & 0.27 & -0.27 & 0.078 \\

\end{tabular}
\end{center}
\end{ruledtabular}
\end{table}
\endgroup
Good agreement is obtained in comparison of the form factors for
group of single-particle-hole states with the experimental data. The
amplitudes of the transition to the positive-parity states
considered in this work have to be reduced by a factor of 2 to
describe the low $q$ data. The form factors for the odd-parity
particle-hole states are in reasonable agreement with the experiment
and need a reduction factor of 1.16 in the amplitudes $\chi$ to
yield excellent agreement. Higher configurations are necessary to
improve the agreement with the low $q$ values of the form factors.


\end{document}